\documentclass[aps,prx,reprint,superscriptaddress,nofootinbib,floatfix]{revtex4-2}

\usepackage{amsmath,amssymb,bm,mathtools}
\usepackage{booktabs}
\usepackage{array}[=2016-10-06] 
\usepackage{graphicx}
\usepackage{xcolor}
\usepackage{enumitem}
\usepackage[hidelinks]{hyperref}
\usepackage{microtype}
\usepackage{placeins}

\newcommand{\Nc}{N_{\mathrm c}}
\newcommand{\Sa}{S_{\alpha}}
\newcommand{\Sb}{S_{\beta}}
\newcommand{\Tr}{\operatorname{Tr}}
\newcommand{\adj}{\operatorname{adj}}

\newcommand{\cO}{\mathcal O}
\newcommand{\cS}{\mathcal S}
\newcommand{\cC}{\mathcal C}

\newcommand{\vct}[1]{\bm{#1}}

\newcommand{\mytitle}{
Differentiated Determinant Graphs Enable Large-Scale Multideterminant Auxiliary-Field Quantum Monte Carlo
}

\begin{document}

\title{\mytitle
}

\author{Jacob Z. Williams}
\affiliation{Theoretical Division, Los Alamos National Laboratory, Los Alamos, New Mexico 87545, USA}

\author{Yu Zhang}
\email{zhy@lanl.gov}
\affiliation{Theoretical Division, Los Alamos National Laboratory, Los Alamos, New Mexico 87545, USA}


\begin{abstract}
Multideterminant trial wave functions can reduce phaseless bias in auxiliary-field quantum Monte Carlo (AFQMC), but the repeated evaluation of their overlap, force bias, and local energy restricts systematic trial improvement.
In this work, we organize particle--hole excitation determinants as a shared arithmetic directed acyclic graph of canonical minors. A coefficient-weighted reverse pass and batched directional passes evaluate the complete estimator without determinant-resolved cofactor tensors while preserving the sparse $\alpha$--$\beta$ coefficient coupling exactly. Controlled complex-valued tests verify the complete kernel to machine precision and show orders-of-magnitude acceleration over direct determinant summation. In molecular calculations, the observed cost remains sublinear through million-configuration selected-CI trials and the same estimator remains usable at one hundred million configurations, where the eventual exact coefficient-contraction boundary becomes visible. For Cr$_2$, an orbital-optimized CAS(12e,42o) trial in a double-$\zeta$ basis yields a potential-energy curve (PEC) and spectroscopic constants in close agreement with experiment. Differentiated determinant graphs therefore convert very large correlated trials from a formal accuracy hierarchy into a practical route for systematic AFQMC studies of strongly correlated molecules.
\end{abstract}

\maketitle

\section{Introduction}

Auxiliary-field quantum Monte Carlo (AFQMC) combines a low-polynomial-scaling stochastic projection with an importance-sampling constraint supplied by a trial wave function~\cite{Hubbard1959, Blankenbecler1981, Hirsch1983, ZhangCarlsonGubernatis1997,Carlson1999, ZhangKrakauer2003, AlSaidi2006, MottaZhang2018, MottaSymmetries2019, MaloneCrossValidation2020, LeePhamReichman2022, Zhang:2026aa}. A single Slater determinant is inexpensive and often accurate, but bond breaking, transition-metal compounds, heavy-element chemistry, and other multireference problems can require a correlated trial wavefunction~\cite{AlSaidiTiOMnO2006,AlSaidiBond2007,PurwantoF22008,PurwantoCr22015,SheeTransitionMetals2019,MahajanSharma2021}.
Complete-active-space and selected-configuration-interaction (sCI) expansions provide a systematic route to such trials~\cite{Roos1980,KnowlesWerner1985,Huron1973,Evangelisti1983,Tubman2016,Holmes2016,Sharma2017,Garniron2019}, and modern implementations have demonstrated expansions containing $10^5$--$10^6$ configurations~\cite{Holmes2016,Tubman2016,Sharma2017,MahajanLeeSharma2022,Malone2023,MahajanBeyond2025}. Their variational energies and phaseless constraints can nevertheless converge slowly with the number of retained configurations, making long expansions necessary in difficult cases. The trial--walker overlap, force bias, and local energy are then evaluated for every walker and imaginary-time step, so even a modest per-configuration overhead is multiplied many times.

The na\"ive multideterminant algorithm treats every full $\alpha$--beta product independently. It factors two occupied-overlap matrices, constructs determinant-specific transition Green's functions, and performs determinant-specific Hamiltonian contractions for each of the $\Nc$ retained configurations~\cite{Suewattana2007,Shee2018,MaloneGPU2020,Malone2023}.
Batched Sherman--Morrison--Woodbury (SMW) updates and GPU kernels reduce the prefactor~\cite{ShermanMorrison1950, Woodbury1950, Hager1989, NukalaKent2009, Shee2018, MaloneGPU2020},
while the unique-spin determinant tree removes repeated spin strings and connects nearby occupied matrices by low-rank updates~\cite{ShiZhang2021}. 
Related real-space QMC algorithms use spin-factorized determinant lists~\cite{ScemamaLargeMSD2016} and table-based low-rank updates to evaluate large multideterminant trial wave functions and their local properties~\cite{Slootman:2026aa}.

A distinct development uses the generalized Wick theorem (GWT)~\cite{Wick1950,Lowdin1955,Thouless1960,OnishiYoshida1966,BalianBrezin1969,Robledo2009,Burton2021}. Every orthogonal CI determinant is represented as a product of particle--hole excitations relative to one reference determinant. Its overlap ratio with a walker becomes a small excitation determinant. First (second) cofactors, signed determinants obtained after deleting one row and one column (two rows and two columns), encode the one- and two-body Hamiltonian insertions. This organization prevents large system-size contractions from being repeated for every configuration~\cite{MahajanSharma2020,MahajanSharma2021} and has been adapted to the ph-AFQMC force bias and local energy~\cite{MahajanLeeSharma2022,MahajanResponse2023}. Representative force-bias costs then change from forms in which $\Nc$ multiplies large orbital contractions to $O(\Nc+XM^2)$, where $M$ is the orbital count and $X$ is the number of Cholesky tensors.

However, GWT still repeats much of the small-matrix algebra: a rank-$k$ excitation requires its determinant, $k^2$ first cofactors, and up to $k^2(k-1)^2$ second cofactors. In this work, we merge identical minors from different spin strings and derivative requests in a directed acyclic graph (DAG). An explicit-cofactor DAG isolates the benefit of sharing common subexpressions. Our differentiated DAG goes further: after unique-spin factorization, a differentiated overlap graph retains only the requested excitation determinants and the recursively shared minors needed to evaluate them. One coefficient-weighted reverse traversal gives the aggregate first derivative, and batched forward directional traversals provide the first and second Cholesky responses. This eliminates determinant-resolved derivative tensors and excitation-matrix inverses. To our knowledge, this is the first use of reverse accumulation and directional second derivatives on a shared determinant-minor graph for the complete multideterminant AFQMC estimator.

We compare na\"ive summation, flat GWT, a unique-spin determinant tree, the explicit-cofactor DAG, and the differentiated DAG on identical complex walkers and Hamiltonians under a common correctness gate. Controlled benchmarks isolate the algebraic gain, while production Cr$_2$ and [Cu$_2$O$_2$]$^{2+}$ calculations test the implementation with physical selected-CI trials containing as many as $10^8$ configurations.

\begin{figure*}[htb]
\centering
\includegraphics[width=\linewidth]{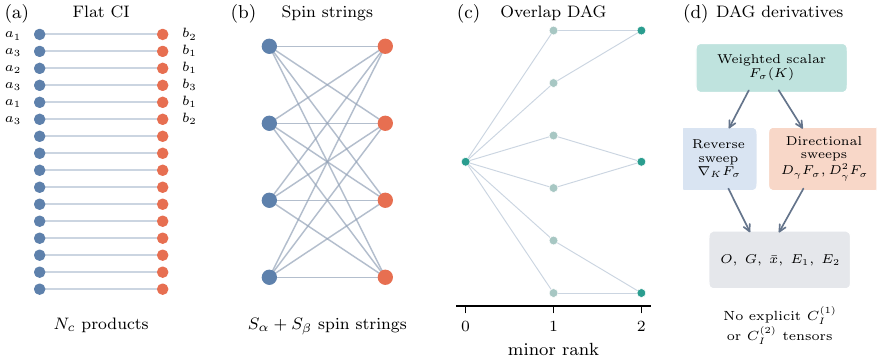}
\caption{
Schematic of the differentiated-DAG algorithm. (a) A na\"ive CI representation repeats $\alpha$ and $\beta$ strings across full configurations. (b) Deduplication exposes $\Sa+\Sb$ unique strings and a sparse bipartite coefficient coupling. (c) The differentiated overlap DAG contains top-level excitation determinants (dark) and recursively shared Laplace dependencies (pale). (d) A reverse traversal of the weighted scalar and batched Cholesky-directional traversals supply the complete AFQMC kernel without explicit determinant-resolved first- or second-cofactor tensors.}
\label{fig:schematic}
\end{figure*}

\section{Theory}

\subsection{AFQMC with a multideterminant trial wavefunction}

Consider a spin-resolved Cholesky representation of the electron-repulsion integrals (ERIs). The electronic Hamiltonian can then be rewritten in the so-called Monte Carlo (MC) form~\cite{BeebeLinderberg1977,Koch2003,Aquilante2007,Motta2019},
\begin{equation}
\hat H =\sum_{\sigma\in\{\alpha,\beta\}}\sum_{pq}
h^\sigma_{pq}\hat a^\dagger_{p\sigma}\hat a_{q\sigma}
+\frac12\sum_{\gamma=1}^{X}\hat v_\gamma^2+E_0,
\label{eq:hamiltonian}
\end{equation}
where any normal-ordering shift is absorbed into $h^\sigma$, $\hat v_\gamma=\sum_{\sigma pq}L^\gamma_{pq} \hat a^\dagger_{p\sigma}\hat a_{q\sigma}$, and ${L^\gamma}$ are the Cholesky tensors. Here, $X$ is the number of retained Cholesky tensors and $E_0$ collects the scalar contributions generated by the chosen rearrangement of the Hamiltonian. This factorized form avoids explicitly storing or repeatedly contracting the full four-index ERI tensor and is the standard starting point for the Hubbard--Stratonovich transformation~\cite{Hubbard1959} and force-bias construction in AFQMC.

The trial wavefunction is an orthogonal multideterminant expansion~\cite{ShiSymmetryProjected2014, ChangTrial2016,VitaliBCS2019,Huggins2022,Amsler2023,JiangMPS2025,Sukurma2025},
\begin{equation}
|\Psi_T\rangle=\sum_{I=1}^{\Nc}c_I|D_I^\alpha D_I^\beta\rangle,
\label{eq:trial}
\end{equation}
where $\lvert D_I^\sigma \rangle$ is a Slater determinant for spin $\sigma$, and a walker remains a product of two Slater determinants, $|\Phi\rangle=|\Phi^\alpha\Phi^\beta\rangle$, with $\Phi^\sigma\in\mathbb C^{M\times N_\sigma}$. Here, $M$ is the number of one-particle orbitals and $N_\sigma$ is the number of electrons with spin $\sigma$. Although the trial state contains $\Nc$ configurations, each walker is represented by only one determinant per spin sector. Consequently, the computational cost associated with the multideterminant trial enters through the repeated evaluation and weighted accumulation of trial--walker overlaps and contractions.

For determinant $I$, the spin-resolved overlap and Green's function are given by
\begin{align}
O_I^\sigma
&=\det \left[(T_I^\sigma)^\dagger\Phi^\sigma\right],\\
G_I^\sigma
&=\Phi^\sigma\left[(T_I^\sigma)^\dagger\Phi^\sigma\right]^{-1}
(T_I^\sigma)^\dagger,
\label{eq:det-green}
\end{align}
where $T_I^\sigma$ contains the occupied trial orbitals of configuration $I$ in spin sector $\sigma$. The matrix $(T_I^\sigma)^\dagger\Phi^\sigma$ is the overlap matrix between the occupied orbitals of trial determinant $I$ and the walker. Its determinant gives their many-electron overlap, while its inverse enters the corresponding one-particle transition density matrix. Thus, $G_I^\sigma$ contains all one-body contractions evaluated between the bra determinant $\langle D_I^\sigma|$ and the walker $|\Phi^\sigma\rangle$, normalized by their spin-resolved overlap.

The unnormalized configuration weight and full trial--walker overlap are
\begin{equation}
z_I=c_I^*O_I^\alpha O_I^\beta,
\quad O=\langle\Psi_T|\Phi\rangle=\sum_I z_I.
\label{eq:weights}
\end{equation}
Each $z_I$ combines the CI coefficient with the overlaps from both spin sectors and may in general be complex. These weights determine the contribution of every trial configuration to all mixed-estimator quantities. The mixed one-body Green's function and force-bias contraction are
\begin{equation}
G^\sigma=\frac{1}{O}\sum_I z_I G_I^\sigma,
\quad
\bar x_\gamma=-\sqrt{\Delta\tau}
\Tr [L^\gamma \sum_\sigma G^\sigma].
\label{eq:forcebias}
\end{equation}
The first expression is a normalized weighted average because a one-body operator depends linearly on the transition Green's function. The force bias is therefore obtained directly from the total mixed Green's function and couples the walker to each Cholesky field through the corresponding one-body contraction.

For a determinant $I$, define
\begin{align}
t_{I\gamma}^\sigma&=\Tr(L^\gamma G_I^\sigma),\\
x_{I\gamma}^\sigma&=\Tr(L^\gamma G_I^\sigma L^\gamma G_I^\sigma).
\end{align}
The quantity $t_{I\gamma}^\sigma$ is the one-body contraction of Cholesky vector $\gamma$ with the determinant-resolved Green's function, whereas $x_{I\gamma}^\sigma$ is the associated exchange-like contraction. The two-body energy of determinant $I$ is then given by
\begin{equation}
E_{2,I}=\frac12\sum_\gamma
\left[\left(t_{I\gamma}^\alpha+t_{I\gamma}^\beta\right)^2 -x_{I\gamma}^\alpha-x_{I\gamma}^\beta\right],
\label{eq:det-e2}
\end{equation}
where the squared term is the direct contribution, which sums over both spins, while the two $x_{I\gamma}^\sigma$ terms are the same-spin exchange contributions. This expression is evaluated separately for each trial determinant because all of its contractions depend on the determinant-specific transition Green's functions.

The MSD local-energy component is then obtained from the determinant-resolved energy as
\begin{equation}
E_2=\frac{1}{O}\sum_I z_I E_{2,I}.
\label{eq:msd-e2}
\end{equation}
In other words, the determinant two-body energies must first be formed using their respective transition Green's functions and only then combined using the configuration weights. This ordering differs fundamentally from the one-body case. Because Eq.~\eqref{eq:det-e2} is quadratic in $G_I$, generally
\begin{equation}
E_2[G^\alpha,G^\beta]
\neq \frac{1}{O}\sum_Iz_IE_2[G_I^\alpha,G_I^\beta].
\label{eq:averaged-g-wrong}
\end{equation}
Evaluating the quadratic functional using the averaged Green's function would introduce spurious cross terms between distinct trial configurations and would therefore not reproduce the multideterminant mixed estimator. Preserving the determinant-resolved structure of the two-body numerator is consequently a central correctness condition for every accelerated backend.

A na\"ive implementation evaluates the occupied-overlap factorizations, transition Green's functions, and Hamiltonian contractions independently for all $\Nc$ configurations at every walker and time step. Its cost is therefore at least linear in $\Nc$ and does not exploit shared spin strings or particle--hole substructures.
A self-contained derivation of this determinant-resolved multideterminant estimator, including the failure of the averaged-Green's-function shortcut in Eq.~\eqref{eq:averaged-g-wrong}, is given in Sec.~\ref{sec:si-native} of the Supplemental Material (SM).

\subsection{Reference formulation and unique-spin reduction}
\label{sec:reference_ratio}

We represent every trial determinant as a particle--hole excitation from a common reference determinant~\cite{Thouless1960, MahajanSharma2020, MahajanSharma2021}. This reference-based construction replaces repeated full occupied-overlap factorizations by one reference factorization and a collection of small excitation determinants. The overlap, density-matrix, and Hamiltonian contractions required by the GWT are obtained from these determinants and their derivatives~\cite{Wick1950,Lowdin1955,Burton2021,MahajanLeeSharma2022}.
For spin $\sigma$, let $\cO_0^\sigma$ be the reference occupation and define
\begin{equation}
B_0^\sigma=\Phi^\sigma[\cO_0^\sigma,:], \quad
K^\sigma=\Phi^\sigma(B_0^\sigma)^{-1}.
\label{eq:reference-kernel}
\end{equation}
Only the $N_\sigma\times N_\sigma$ reference matrix is factorized. A spin string $a$ that replaces occupied positions $H_a^\sigma$ by particle orbitals $P_a^\sigma$ has the exact relative overlap
\begin{equation}
r_a^\sigma=s_a^\sigma \det K^\sigma[P_a^\sigma,H_a^\sigma],
\label{eq:overlap-ratio}
\end{equation}
where $s_a^\sigma$ restores the canonical fermionic sign.
Writing the small excitation matrix as $A_a^\sigma=K^\sigma[P_a^\sigma,H_a^\sigma]$, its first and second derivatives are signed deletion minors,
\begin{align}
\frac{\partial\det A}{\partial A_{\mu\nu}}
&=(-1)^{\mu+\nu}\det A[\widehat\mu|\widehat\nu],
\label{eq:first-cofactor}\\
\frac{\partial^2\det A}
{\partial A_{\mu\nu}\partial A_{\rho\lambda}}
&=\epsilon_{\mu\nu\rho\lambda}
\det A[\widehat{\mu\rho}|\widehat{\nu\lambda}].
\label{eq:second-cofactor}
\end{align}
The second derivative vanishes when the two differentiated entries share a row or column. These derivatives generate the one-body and same-spin two-body GWT insertions, respectively, while products of first derivatives generate the opposite-spin contribution. Their polynomial deleted-minor form requires no inverse of $A_a^\sigma$ and therefore remains valid even when an excitation overlap ratio vanishes. The reference projectors, excitation-index conventions, and overlap factorization are detailed in Sec.~\ref{sec:si-reference} of the SM.

Repeated spin strings are evaluated only once. If $a(I)$ and $b(I)$ identify the unique $\alpha$ and $\beta$ strings in configuration $I$, the relative trial--walker overlap and the opposite-spin environment weights are
\begin{align}
\cS&=\sum_{I=1}^{\Nc}c_I^*r_{a(I)}^\alpha r_{b(I)}^\beta,\nonumber\\
q_a^\alpha&=\sum_{I:a(I)=a}c_I^*r_{b(I)}^\beta,
&q_b^\beta&=\sum_{I:b(I)=b}c_I^*r_{a(I)}^\alpha.
\label{eq:environment-weights}
\end{align}
Thus $\cS=\sum_aq_a^\alpha r_a^\alpha=\sum_bq_b^\beta r_b^\beta$.
All terms containing a nontrivial feature in only one spin sector reduce to these environment sums. Terms coupling nontrivial features from both spins retain the original sparse CI coefficients exactly. The complete unique-spin reduction, including the irreducible opposite-spin coefficient coupling, is given in Sec.~\ref{sec:si-environments} of the SM.

For example, embedding the environment-weighted first cofactors in the common reference-hole--particle space gives a compact matrix $\cC^\sigma$. If $G_0^\sigma$ is the reference transition density and $Q_0^\sigma=I-G_0^\sigma$, the complete mixed density is
\begin{equation}
G^\sigma=G_0^\sigma+\frac{K^\sigma\cC^\sigma Q_0^\sigma}{\cS}.
\label{eq:gwt-green}
\end{equation}
Thus the CI reduction is performed in the small excitation space before the full-$M$ matrix products. The cofactor conventions and the derivations of Eq.~\eqref{eq:gwt-green} and the corresponding one- and two-body estimator identities are given in Secs.~\ref{sec:si-cofactors}--\ref{sec:si-two-body} of the SM.

\subsection{Shared determinant graph and differentiated evaluation}

The excitation determinants are minors of $K^\sigma$. For canonically ordered row and column tuples $R$ and $C$, the first-row Laplace recurrence
\begin{equation}
M(R,C)=\sum_{j=1}^{|C|}(-1)^{j-1}K_{r_1c_j}M(R\setminus r_1,C\setminus c_j)
\label{eq:laplacedag}
\end{equation}
defines an arithmetic DAG when every pair $(R,C)$ is stored once. Identical minors are therefore shared across spin strings and derivative paths. The explicit-cofactor DAG includes the single- and double-deletion minors requested by the GWT, whereas the differentiated overlap DAG contains only the overlap roots and the dependencies generated by Eq.~\eqref{eq:laplacedag}.

The graph topology depends only on the excitation patterns and is compiled once with canonical $(R,C)$ tuples as hash keys. At each walker evaluation, only the kernel entries and node values change. Rank ordering supplies a topological forward schedule, so a minor shared by several configurations is evaluated once and reused by every parent. The explicit-cofactor construction separates the gain from common-subexpression sharing: it still recovers a first-cofactor matrix and second-cofactor tensor for each unique spin string. The differentiated overlap graph removes those derivative-only roots and differentiates the overlap circuit itself. This also distinguishes it from a determinant tree, which assigns one occupied-matrix parent to each spin string and propagates full determinant and inverse information along that edge. A minor-DAG node can instead serve arbitrarily many overlap roots and derivative paths.

For either spin sector, we form the environment-weighted scalar
\begin{equation}
F^\sigma(K^\sigma)=\sum_{a=1}^{S_\sigma}q_a^\sigma s_a^\sigma
M^\sigma(P_a^\sigma,H_a^\sigma)=\cS.
\label{eq:weighted-scalar}
\end{equation}
Let $v_n$ denote a node value and write a graph edge as $e=(n\rightarrow m)$, with Laplace sign $s_e$ and associated kernel element $(r_e,c_e)$. The node recurrence is
\begin{equation}
v_n=\sum_{e=(n\rightarrow m)}s_eK^\sigma_{r_ec_e}v_m.
\label{eq:node-recurrence}
\end{equation}
After seeding each overlap root by
$\bar v_{n(a)}\mathrel{+}=q_a^\sigma s_a^\sigma$, one reverse traversal applies
\begin{align}
\Gamma^\sigma_{r_ec_e}&\mathrel{+}=\bar v_ns_ev_m,\nonumber\\
\bar v_m&\mathrel{+}=\bar v_ns_eK^\sigma_{r_ec_e},
\label{eq:reverse-dag}
\end{align}
where $\bar v_n=\partial F^\sigma/\partial v_n$ and $\Gamma^\sigma=\partial F^\sigma/\partial K^\sigma$. Adjoint contributions from all parents are summed, so a shared minor is differentiated only once. Moreover, $\cC^\sigma=(\Gamma^\sigma)^T$. The same aggregate gradient supplies the mixed density, force bias, one-body energy, and every first-connected contraction.

The two-body estimator needs selected Hessian contractions rather than the full Hessian. For a Cholesky-induced direction $B^{\sigma\gamma}$, introduce $K(t)=K+tB^{\sigma\gamma}$. Differentiating Eq.~\eqref{eq:node-recurrence} gives the node tangents
\begin{align}
\dot v_n^\gamma&=\sum_{e=(n\rightarrow m)}s_e
\left(B^{\sigma\gamma}_{r_ec_e}v_m
+K^\sigma_{r_ec_e}\dot v_m^\gamma\right),
\label{eq:first-tangent}\\
\ddot v_n^\gamma&=\sum_{e=(n\rightarrow m)}s_e
\left(2B^{\sigma\gamma}_{r_ec_e}\dot v_m^\gamma
+K^\sigma_{r_ec_e}\ddot v_m^\gamma\right).
\label{eq:second-tangent}
\end{align}
At root $a$, $s_a^\sigma\dot v_a^\gamma$ is the first-connected Cholesky feature $f_{a\gamma}^\sigma$, while the weighted same-spin double insertion is $\frac12\sum_{a\gamma}q_a^\sigma s_a^\sigma\ddot v_a^\gamma$. Processing the $X$ directions in blocks of width $b$ limits tangent storage to $O(bV_{\rm diff})$. Both traversals act directly on polynomial recurrences and remain well defined for singular excitation matrices. Figure~\ref{fig:schematic} summarizes this sequence. A one-to-one correspondence with the explicit GWT cofactors is given in Sec.~\ref{sec:si-dag-equivalence} and Table~\ref{tab:si-correspondence} of the SM.

\subsection{Energy estimators}

The graph derivatives are contracted with full-system quantities constructed only for the reference determinant. For each spin and walker, define
\begin{align}
H^\sigma&=Q_0^\sigma h^\sigma K^\sigma,\nonumber\\
B^{\sigma\gamma}&=Q_0^\sigma L^\gamma K^\sigma,\nonumber\\
C^{\sigma\gamma}&=Q_0^\sigma L^\gamma G_0^\sigma L^\gamma K^\sigma.
\label{eq:reference-intermediates}
\end{align}
Here $H^\sigma$ supplies the one-body insertion, $B^{\sigma\gamma}$ the direct Cholesky response, and $C^{\sigma\gamma}$ the reference-dressed same-spin exchange response. If $e_{1a}^\sigma$, $f_{a\gamma}^\sigma$, and $g_{a\gamma}^\sigma$ denote the corresponding first-cofactor features, reverse accumulation gives their environment-weighted sums directly:
\begin{align}
\sum_aq_a^\sigma e_{1a}^\sigma&=\Gamma^\sigma:H^\sigma,\nonumber\\
\sum_aq_a^\sigma f_{a\gamma}^\sigma&=\Gamma^\sigma:B^{\sigma\gamma},\nonumber\\
\sum_aq_a^\sigma g_{a\gamma}^\sigma&=\Gamma^\sigma:C^{\sigma\gamma}.
\label{eq:aggregate-first-contractions}
\end{align}
The colon is an algebraic bilinear contraction without complex conjugation. The weighted same-spin second-cofactor feature $d_a^\sigma$ is obtained from the second root tangents above, so neither derivative tensor is materialized.

Let $E_{1,0}$ and $E_{2,0}$ be the reference one- and two-body energies and set $T_{0\gamma}=\Tr[L^\gamma(G_0^\alpha+G_0^\beta)]$. Defining
\begin{equation}
u_a^\sigma=\sum_\gamma \left(T_{0\gamma}f_{a\gamma}^\sigma-g_{a\gamma}^\sigma\right) +d_a^\sigma,
\end{equation}
the exact multideterminant contributions are
\begin{align}
E_1=&E_{1,0}+\frac{1}{\cS}
\left(\sum_aq_a^\alpha e_{1a}^\alpha
+\sum_bq_b^\beta e_{1b}^\beta\right),
\label{eq:gwt-e1}\\
E_2=&E_{2,0}+\frac{1}{\cS}\sum_Ic_I^*\bigg[
r_{b(I)}^\beta u_{a(I)}^\alpha+r_{a(I)}^\alpha u_{b(I)}^\beta
\nonumber\\
&\hspace{20mm}+\sum_\gamma f_{a(I)\gamma}^\alpha f_{b(I)\gamma}^\beta\bigg].
\label{eq:gwt-e2}
\end{align}
The first two corrections in Eq.~\eqref{eq:gwt-e2} reduce through the environment weights. The last, opposite-spin double-connected contribution retains the exact sparse contraction
\begin{equation}
\sum_{I\gamma}c_I^*f_{a(I)\gamma}^\alpha f_{b(I)\gamma}^\beta.
\label{eq:exact-cross-term}
\end{equation}
Which is essential: applying the single-determinant quadratic energy functional to the CI-averaged Green's function would instead introduce spurious cross-configuration terms. Equations~\eqref{eq:gwt-e1} and \eqref{eq:gwt-e2} therefore retain the complete determinant-resolved estimator while eliminating determinant-specific full Green's matrices and cofactor tensors. Detailed index-level identities and the complete flat-GWT assembly are collected in Secs.~\ref{sec:si-one-body}--\ref{sec:si-flat-kernel} of the SM.

\subsection{Cost model and exactness boundary}

Let $V_{\rm diff}$ and $E_{\rm diff}$ denote the total numbers of nodes and edges, respectively, in the two spin-sector differentiated overlap DAGs. For each walker, the forward value traversal and reverse gradient traversal both cost $O(E_{\rm diff})$. Propagating all $X$ Cholesky directions costs $O(XE_{\rm diff})$, with the direction-independent node values reused across direction blocks. The corresponding tangent memory is $O(bV_{\rm diff})$ for block width $b$. A schematic per-walker cost is
\begin{align}
T_{\rm diffDAG}=&T_{\rm ref}(M,N_\sigma,X) +O(E_{\rm diff})+O(XE_{\rm diff})
\nonumber\\
& +T_{\rm agg}(M,N_\sigma,X) +O(\Nc X),
\label{eq:cost}
\end{align}
where $T_{\rm ref}$ and $T_{\rm agg}$ contain the determinant-independent reference and full-system contractions. The final term is Eq.~\eqref{eq:exact-cross-term}. Graph differentiation compresses repeated minor algebra but cannot compress arbitrary independent CI coefficients, so $O(\Nc X)$ remains the exact asymptotic boundary. Observed sublinear timing over a finite range is therefore a preasymptotic reduction of the dominant work, not a claim of sublinear asymptotic complexity.

To benchmark the differentiated DAG, we implemented five exact MSD-AFQMC backends: na\"ive determinant summation, flat generalized Wick theorem (GWT), the unique-spin determinant tree of Shi and Zhang~\cite{ShiZhang2021}, an explicit-cofactor DAG, and the differentiated DAG. These backends differ in where they repeat work. Na\"ive summation performs two occupied-overlap factorizations and a full energy evaluation for every CI configuration. Flat GWT reduces the system-scale work to the two reference spin determinants but still constructs the small determinant and cofactor objects repeatedly. The unique-spin tree reduces this repetition to the $S_\alpha+S_\beta$ distinct strings through low-rank occupied-matrix updates. The explicit-cofactor DAG shares common deletion minors between those strings, but retains their separate derivative objects. Finally, the differentiated DAG accumulates the required derivatives on the shared differentiated overlap graph and materializes only the top-root first responses needed by Eq.~\eqref{eq:exact-cross-term}. This sequence provides controlled ablations of reference reorganization, spin-string deduplication, minor sharing, and graph differentiation. Operation counts, storage requirements, and a tabular comparison are given in Sec.~\ref{sec:si-cost} of the SM.

\section{Results and Discussion}
\label{sec:results}

\subsection{Controlled benchmarks of the complete AFQMC kernel}
\label{sec:numerical-equivalence}

The results in this subsection are obtained entirely with a common Python/NumPy proof-of-principle prototype of the five backends. They test the backends' algebraic organization on synthetic data and are distinct from the compiled production differentiated-DAG calculations reported below. The controlled benchmark uses dense complex data with $M=40$ spatial orbitals, $N_\alpha=N_\beta=10$, $X=100$ Cholesky tensors, excitation ranks through four, and balanced unique-spin pools. Determinant counts range from 30 to $10^4$. Each method evaluates the same overlap, mixed Green's functions, force bias, and one- and two-body local energies. Details of the deterministic construction, numerical validation, timing table, and graph-structure profiles are given in the SM.

Every accelerated backend agrees with na\"ive determinant summation to better than $1.2\times10^{-14}$ across the full benchmark. The SI also shows that applying the single-determinant quadratic energy functional to the CI-averaged Green's function can produce a relative error as large as 0.334, confirming the need for the connected first- and second-derivative terms retained here.

\begin{figure}[htb]
\centering
\includegraphics[width=\linewidth]{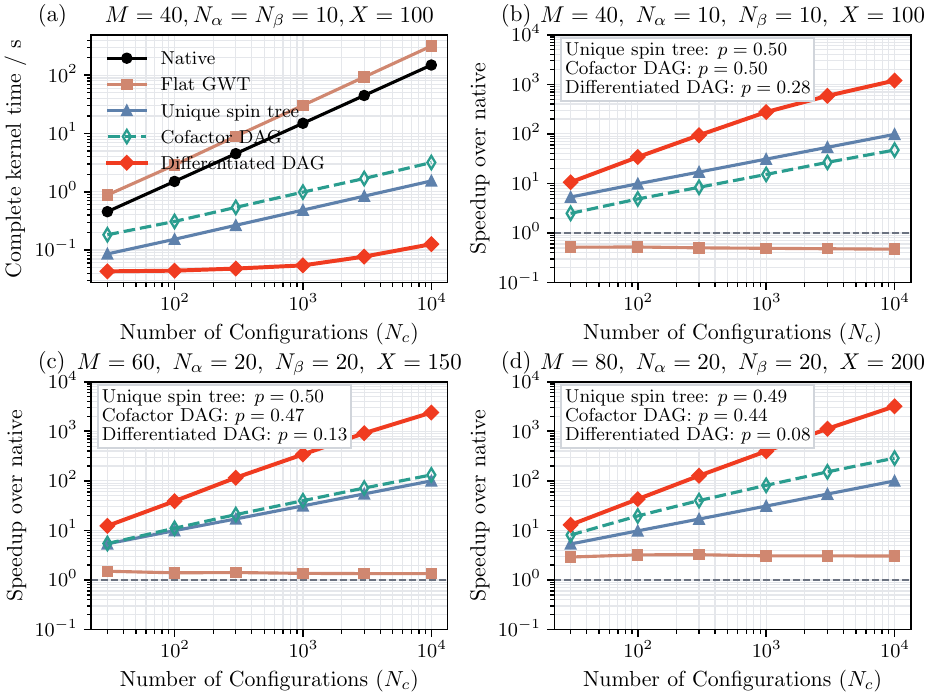}
\caption{Controlled-benchmark timing. (a) Complete-kernel wall time and (b) speedup over na\"ive summation for $(M,N_\alpha,N_\beta,X)=(40,10,10,100)$. (c,d) Speedups for $(60,20,20,150)$ and $(80,20,20,200)$, respectively. Labels give finite-range exponents from $T\propto\Nc^p$.}
\label{fig:scaling}
\end{figure}

Figure~\ref{fig:scaling}(a) shows the complete-kernel wall times, and panel (b) shows the speedup relative to na\"ive determinant summation. At $\Nc=10^4$, the differentiated DAG is therefore 1192 times faster than na\"ive summation, 25.2 times faster than the explicit-cofactor DAG, and 12.2 times faster than the unique-spin tree. The fitted timing exponents over the displayed range are 1.00, 1.02, 0.50, 0.49, and 0.28 in the same order. The explicit-cofactor comparison shows that the additional gain comes from accumulating derivatives through the shared graph, rather than from minor sharing alone.

Panels (c) and (d) repeat the speedup comparison for larger orbital spaces, with $(M,N_\alpha,N_\beta,X)=(60,20,20,150)$ and $(80,20,20,200)$, respectively. The differentiated-DAG timing exponent decreases from 0.28 for $M=40$ to 0.13 and 0.08, while the unique-spin-tree exponents remain near 0.5. Over this finite determinant range, increasing the orbital and Cholesky dimensions increases the determinant-independent reference and Hamiltonian work relative to the incremental graph traversal. Adding configurations therefore contributes an even smaller fraction of the complete-kernel cost in the larger spaces. These exponents remain preasymptotic summaries and do not remove the exact $O(\Nc X)$ boundary.

The graph benefit depends on the number and balance of unique spin strings and on excitation rank. The SI reports these sensitivities and the one-time compilation costs. The exact sparse opposite-spin coefficient contraction remains $O(\Nc X)$, so the exponent in Fig.~\ref{fig:scaling} describes the preasymptotic range rather than a change in asymptotic complexity.

All controlled and molecular scaling measurements reported here were performed with one CPU thread on a MacBook Pro with a 14-core Apple M4 Pro processor and 48~GB of memory. The production timings therefore measure algorithmic scaling on a common workstation platform. The calculations do not use GPU acceleration.

\subsection{Production scaling in real molecules: Cr$_2$ and [Cu$_2$O$_2$]$^{2+}$}
\label{subsec:cr2-production-scaling}

{\bf  Cr$_2$ molecule.} We next test the production differentiated-DAG implementation with physical selected-CI trials for Cr$_2$, a demanding multireference molecule. The calculation uses $R=1.68$~\AA, a spin-free exact-two-component cc-pVDZ-DK Hamiltonian with $M=86$ orbitals and $X=648$ Cholesky tensors, and Dice SHCI expansions in a CAS(28e,42o) active space. The expansions are truncated from $\Nc=10^3$ to $10^8$ configurations. Each timing uses the same ten-walker, 125-step trajectory, with the local energy evaluated every 25 steps. Times are averaged over the five resulting measurement blocks and exclude SHCI generation and one-time trial and graph construction.

\begin{figure}[htb]
  \centering
  \includegraphics[width=\linewidth]{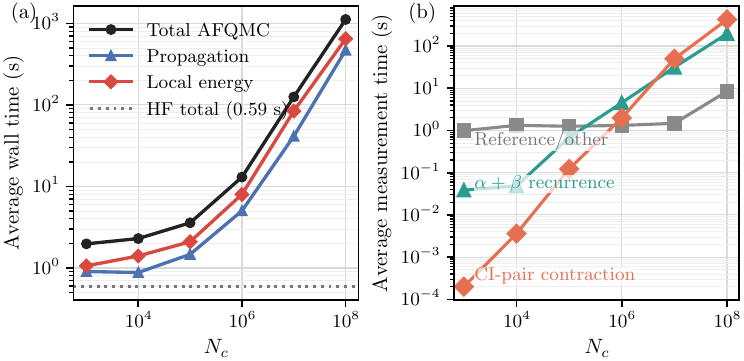}
  \caption{Production differentiated-DAG timing for Cr$_2$ at $R=1.68$~\AA with CAS(28e,42o) SHCI trials. (a) Total, propagation, and local-energy times per 25-step block for ten walkers; the dotted line is the single-determinant total. (b) Local-energy decomposition into the graph recurrence, CI-pair contraction, and remaining work. Each point is one run on one CPU thread.
  }
  \label{fig:scaling_cr2}
\end{figure}

The production calculation retains a weak determinant-count dependence through the million-configuration regime. Increasing the trial from $10^3$ to $10^6$ configurations, a factor of $10^3$ in expansion length, increases the average total time per measurement block from 1.98 to 13.0~s, a factor of 6.59. A log--log fit over this interval gives effective exponents of 0.265 for the total time and 0.281 for the local energy. These are finite-range summaries: the cost contains determinant-independent reference work, graph work governed by the number and topology of unique spin strings, and the exact CI coefficient contraction.

The crossover predicted by the cost analysis becomes visible for the two largest trials. At $10^7$ and $10^8$ configurations, the total times per block are 125 and 1111~s, respectively, and a fit over the full $10^3$--$10^8$ range gives an effective total-time exponent of 0.558. The CI-pair contraction contributes less than 0.3\% of the differentiated local-energy time through $10^4$ configurations, 25.2\% at $10^6$, 60.2\% at $10^7$, and 67.5\% at $10^8$. Its fitted exponent over $10^5$--$10^8$ is 1.20. Thus the large-trial upturn is the expected emergence of the exact $O(\Nc X)$ coefficient-contraction boundary, rather than a loss of graph compression. The calculation nevertheless extends the same production estimator, without changing its algebraic accuracy, from thousands to one hundred million selected configurations.

{\bf [Cu$_2$O$_2$]$^{2+}$ molecule}.
We also benchmark the differentiated-DAG production path for the $f=0$ structure of [Cu$_2$O$_2$]$^{2+}$ in the BS1 basis, matching the molecular system used in the GPU MSD-AFQMC study of Huang \textit{et al.}~\cite{HuangGPU2024}. The correlated space contains 32 electrons in 108 orbitals, with 675 Cholesky tensors. Each point uses ten walkers and 100 propagation steps, with one local-energy measurement every 25 steps. The reported time is the average per 25-step block.

The total block time increases from 3.71~s at $10^3$ configurations to 21.5~s at $10^6$, giving effective exponents of 0.243 for the total calculation and 0.268 for the local energy over the full range. More stringently, the total-time exponent is 0.333 over $10^4$--$10^6$ and 0.527 over the final decade, $10^5$--$10^6$. Thus, even at one million determinants, the differentiated-DAG calculation has not entered a linear regime. At $10^6$ configurations, the exact CI-pair contraction accounts for only 13.9\% of the differentiated local-energy time, so graph and reference-centered work still dominate this range.

For context, Fig.~1(a) of Huang \textit{et al.} reports approximately 700~s on one CPU core at $10^6$ determinants for the same nominal ten-walker, 25-step block~\cite{HuangGPU2024}, compared with 21.5~s on one CPU thread here. This roughly 30-fold difference in reported wall time is only indicative: the hardware, implementations, and propagation kernels are not identical. A more informative qualitative comparison is the determinant-count dependence. The earlier CPU curve becomes approximately linear beyond $10^4$ determinants, whereas the differentiated-DAG calculation remains sublinear through $10^6$. This contrast is consistent with the reduction in effective scaling expected from sharing and differentiating the determinant graph, although a controlled cross-code benchmark would be required to isolate the magnitude of that advantage.

\begin{figure}[htb]
  \centering
  \includegraphics[width=\linewidth]{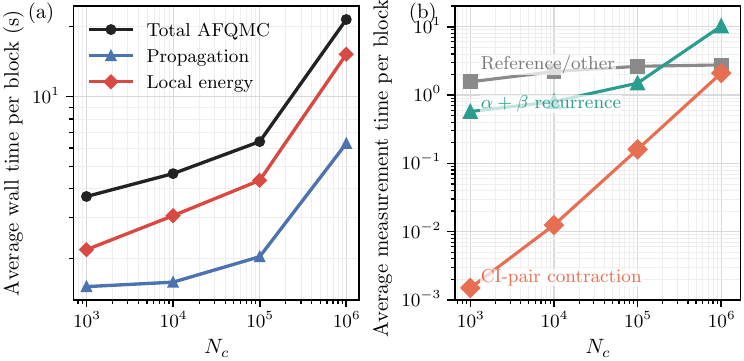}
  \caption{Production differentiated-DAG scaling for [Cu$_2$O$_2$]$^{2+}$ with a CAS(32e,108o) selected-CI trial. Times are averages per 25-step block for ten walkers on one
  CPU thread.}
  \label{fig:scaling_cu2o2}
\end{figure}

\subsection{Cr$_2$ PEC and spectroscopic constants}

\begin{figure*}[t]
  \centering
  \includegraphics[width=\linewidth]{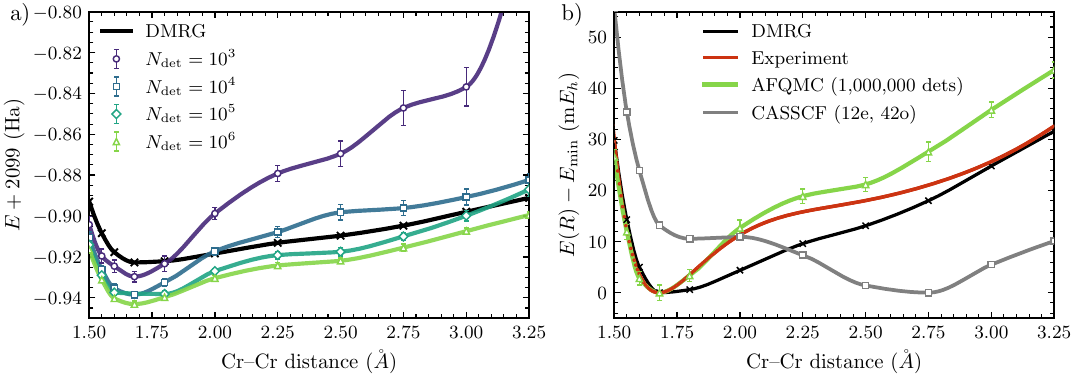}
  \caption{%
  Cr$_2$ PECs from phaseless AFQMC with orbital-optimized CAS(12e,42o) SHCI-CASSCF trials. (a)~Projected energies versus $N_{\mathrm{det}}$ and the DMRG reference. A constant $2099~E_h$ is added. (b)~Minimum-aligned shapes of the $10^6$-configuration AFQMC PEC, its unprojected selected-CI trial, DMRG, and experiment. Error bars are one standard error. Lines are PCHIP interpolations.
  }
  \label{fig:cr2_pec}
\end{figure*}

We next test whether systematically enlarging a selected-CI trial improves the phaseless constraint in a chemically demanding calculation, rather than only providing a favorable scaling benchmark. Figure~\ref{fig:cr2_pec}(a) shows the Cr$_2$ phaseless-AFQMC PECs obtained with increasing SHCI-CASSCF trials in an active space of (12e, 42o), together with the DMRG reference of Larsson \textit{et al.}~\cite{LarssonZhaiUmrigarChan2022} (using an active space of (28e, 76o)), while Fig.~\ref{fig:cr2_pec}(b) compares the relative PEC obtained with the largest trial wave function against the CASSCF (12e, 42o) and DMRG results, as well as the refitted experimental curve of Casey and Leopold~\cite{CaseyLeopold1993} reported in Ref.~\cite{LarssonZhaiUmrigarChan2022}. The CASSCF trials and corresponding energies are computed using Dice~\cite{Sharma:2017aa}. The orbital optimization used a CAS(12e,42o) active space and a convergence tolerance of $10^{-4}$. Retaining CI coefficients with magnitude at least $5\times10^{-5}$ produced between 4 and 8.4 million configurations, depending on the geometry. For the AFQMC calculations, these expansions were truncated to $N_{\mathrm{det}}=10^3$, $10^4$, $10^5$, or $10^6$ configurations. Note that the multideterminant trial enters both importance sampling and local-energy evaluation throughout each AFQMC trajectory.

Fig.~\ref{fig:cr2_pec}(a) shows that Increasing the trial size systematically improves the predicted PEC. The largest change occurs between $10^3$ and $10^4$ configurations, but the curve continues to converge as $N_{\mathrm{det}}$ increases to $10^6$, particularly away from equilibrium. This progression demonstrates the importance of large trial expansions for controlling the geometry-dependent phaseless bias in Cr$_2$.

Orbital optimization is also essential. The orbital-optimized $10^6$-configuration CAS(12e,42o) trial gives a minimum-aligned RMS deviation of $5.7~\mathrm{m}E_h$ from experiment over $1.50$--$3.25$~\AA, whereas the corresponding trial constructed with unoptimized CASCI orbitals gives $19.5~\mathrm{m}E_h$. The CASCI data are reported in Sec.~\ref{sec:si-cr2-casci} and Fig.~\ref{fig:cr2_pec_casci} of the SI. Both RMS values use shape-preserving interpolation on a uniform grid after independently aligning each minimum.

The million-configuration AFQMC curve closely follows experiment through $R=2.5$~\AA, with the remaining visible deviation concentrated at the longest bonds, where the double-$\zeta$ curve rises somewhat too steeply. Earlier CASSCF and DMRG calculations found a substantial geometry-dependent basis correction and used TZ--QZ extrapolation, checked against 5Z, to obtain a complete-basis-set PEC in close agreement with experiment~\cite{PurwantoCr22015}. This supports the expectation that increasing the present cc-pVDZ-DK basis will lower the stretched-bond shoulder and improve the residual discrepancy, although a direct basis-set sequence is required to quantify that change.

To extract the spectroscopic constants in Table~\ref{tab:morse}, we fit the AFQMC PEC to the expanded Morse oscillator used by Larsson \textit{et al.}~\cite{LarssonZhaiUmrigarChan2022}. The dissociation asymptote was fixed by an independent Cr-atom AFQMC calculation with the corresponding CAS(6e,21o), the same cc-pVDZ-DK Hamiltonian, CASSCF and coefficient thresholds, walker protocol, and estimator settings as the dimer. The resulting atomic energy, $-1049.94259\pm 0.00042~E_h$, defines $D_e=2E_{\rm Cr}-E_{\rm min}$. The harmonic constant $\omega_e$ is the linear coefficient of a second-order Dunham fit to the lowest nine $J=0$ vibrational levels of the fitted potential for Cr$_2$.

\begin{table}[tb]
\caption{Equilibrium distance $R_e$, dissociation energy $D_e$, and harmonic vibrational constant $\omega_e$ obtained from expanded-Morse-oscillator fits. The AFQMC and DMRG (extrapolated at infinite bond dimension limit) calculations use the cc-pVDZ-DK basis.}
\label{tab:morse}
\centering
\footnotesize
\begin{tabular}{
  >{\raggedright\arraybackslash}p{0.34\linewidth}
  >{\raggedright\arraybackslash}p{0.2\linewidth}
  >{\raggedright\arraybackslash}p{0.23\linewidth}
  >{\raggedright\arraybackslash}p{0.17\linewidth}
  }
\hline\hline
Method/CAS & $R_e~(\text{\AA})$ & $D_e~(\mathrm{eV})$ & $\omega_e~(\mathrm{cm}^{-1})$ \\
\hline
DMRG (28e,76o)~\cite{LarssonZhaiUmrigarChan2022} & 1.7292 & 1.5814 & 324.6 \\  %
AFQMC (12e,42o) & 1.6819 & 1.5861  & 489.2\\
\hline
Experiment & 1.6788~\cite{BondybeyEnglish1983} & \mbox{$1.56\pm0.06$}~\cite{Simard1998} & 480.6~\cite{CaseyLeopold1993} \\
\hline
\end{tabular}
\end{table}

As shown in Table~\ref{tab:morse}, the AFQMC constants (with a CAS(12e,42o) trial) are in quantitative agreement with experiment: $R_e$ differs by $0.0031$~\AA, $D_e$ by $0.026$~eV and lies within the experimental uncertainty, and $\omega_e$ differs by $8.6~\mathrm{cm}^{-1}$. Thus, the orbital-optimized trial recovers both the near-equilibrium curvature and the binding scale despite the modest basis, while the residual limitation is localized primarily in the long-bond shape.

\section{Conclusions}

We introduced a numerically exact differentiated arithmetic-DAG formulation of the generalized-Wick kernels required by multideterminant AFQMC. After factoring the trial into unique $\alpha$ and $\beta$ strings, the method compiles only the shared overlap-minor recurrence. A coefficient-weighted reverse traversal gives the aggregate first derivative, while batched Cholesky-directional traversals give the first and second responses entering the two-body local energy. This organization eliminates explicit per-string $k^2$ first-cofactor matrices and $k^4$ second-cofactor tensors without constructing a full Green's matrix for every CI determinant.

Across the controlled tests, the complete complex-valued implementation agrees with na\"ive determinant summation to machine precision.
For [Cu$_2$O$_2$]$^{2+}$ at $10^6$ configurations, an indicative cross-code comparison gives a substantially shorter one-thread CPU time than a previously reported one-core result. More importantly, unlike that earlier CPU curve, the differentiated-DAG timing has not reached linear determinant-count scaling.
For the Cr$_2$ PEC, an orbital-optimized, one-million-configuration CAS(12e,42o) trial lowers the minimum-aligned shape deviation from experiment to $5.7~\mathrm{m}E_h$ and gives $R_e$, $D_e$, and $\omega_e$ in quantitative agreement with experiment. The systematic improvement from $10^3$ to $10^6$ configurations directly demonstrates the accuracy gained by enlarging the trial. 

Although demonstrated here in phaseless AFQMC, the differentiated kernels are not tied to the phaseless constraint. They reorganize the overlap and local-energy algebra and therefore apply directly to free-projection AFQMC and Diffusion Monte Carlo~\cite{MaloneCrossValidation2020} with multideterminant trials.
Further gains could target the remaining coefficient contraction through controlled low-rank factorizations, hierarchical sparsity, or stochastic treatment of the selected-CI tail. Such extensions would introduce an approximation or sampling error that can be quantified separately from the exact graph algorithm developed here. By reducing the deterministic cost of long trials, the differentiated DAG makes systematic trial improvement practical for AFQMC studies of strongly correlated molecules and other multireference systems.

More broadly, the construction accelerates determinant overlaps and low-order transition-density derivatives rather than a feature specific to imaginary-time propagation. Analogous differentiated graphs may therefore benefit other generalized-Wick frameworks, including symmetry-projected wave functions and nonorthogonal configuration-interaction methods~\cite{ShiSymmetryProjected2014,Robledo2009,Burton2021}. Adapting the graph structure to these state representations is a direction for future work.

\begin{acknowledgments}
We acknowledge support from the US DOE, Office of Science, Basic Energy Sciences, Chemical Sciences, Geosciences, and Biosciences Division under Triad National Security, LLC (``Triad'') contract Grant 89233218CNA000001 (FWP: LANLECF7).
This research used computational resources provided by the Institutional Computing (IC) Program and the Darwin testbed at Los Alamos National Laboratory (LANL), funded by the Computational Systems and Software Environments subprogram of LANL's Advanced Simulation and Computing program.
LANL is operated by Triad National Security, LLC, for the National Nuclear Security Administration of the US Department of Energy (Contract No. 89233218CNA000001).
\end{acknowledgments}

\bibliography{references}

\begin{widetext}
  \appendix

  \clearpage
  
\begin{center}
 \Large\bf Supplementary Materials for ``\mytitle"
\end{center}

\renewcommand{\thefigure}{S\arabic{figure}}
\setcounter{figure}{0}

The Supplementary Materials (SM) give a self-contained derivation of the flat generalized Wick theorem (GWT) baseline used in the main text. They specify how the multideterminant overlap, mixed one-particle density matrices, force bias, and complete one- and two-body local energies follow from particle--hole excitation determinants and their derivatives. They also give the graph recurrences, backend comparison, numerical safeguards, controlled benchmark construction, and extended results moved from the main text. The flat GWT, explicit-cofactor-DAG, and differentiated-DAG backends evaluate the same polynomial quantities and differ only in how repeated determinant and derivative work is organized.
A colon denotes a bilinear contraction without complex conjugation, $A:B=\sum_{ij}A_{ij}B_{ij}$. Complex conjugation is applied to the CI coefficients because the trial state appears as a bra.

\section{Determinant-resolved multideterminant AFQMC estimator}
\label{sec:si-native}

Consider the Cholesky-factorized Hamiltonian
\begin{equation}
\hat H
=\sum_{\sigma\in\{\alpha,\beta\}}\sum_{pq}
h_{pq}^{\sigma}\hat a_{p\sigma}^{\dagger}\hat a_{q\sigma}
+\frac{1}{2}\sum_{\gamma=1}^{X}\hat v_{\gamma}^{2}+E_0,
\quad
\hat v_{\gamma}=\sum_{\sigma pq}L_{pq}^{\gamma}
\hat a_{p\sigma}^{\dagger}\hat a_{q\sigma},
\label{eq:si-hamiltonian}
\end{equation}
where any one-body term generated by the chosen normal-ordering convention is
included in $h^{\sigma}$. The orthogonal multideterminant trial state and a
walker are
\begin{equation}
|\Psi_T\rangle=\sum_{I=1}^{\Nc}c_I
|D_I^{\alpha}D_I^{\beta}\rangle,
\quad
|\Phi\rangle=|\Phi^{\alpha}\Phi^{\beta}\rangle,
\label{eq:si-trial-walker}
\end{equation}
with $\Phi^{\sigma}\in\mathbb C^{M\times N_{\sigma}}$. For a spin
determinant whose occupied-orbital coefficient matrix is $T_I^{\sigma}$,
define
\begin{align}
O_I^{\sigma}
&=\det\!\left[(T_I^{\sigma})^{\dagger}\Phi^{\sigma}\right],\\
G_I^{\sigma}
&=\Phi^{\sigma}
\left[(T_I^{\sigma})^{\dagger}\Phi^{\sigma}\right]^{-1}
(T_I^{\sigma})^{\dagger}.
\label{eq:si-native-green}
\end{align}
The unnormalized configuration weight and total overlap are
\begin{equation}
z_I=c_I^*O_I^{\alpha}O_I^{\beta},
\quad
O=\langle\Psi_T|\Phi\rangle=\sum_Iz_I.
\label{eq:si-native-weight}
\end{equation}
Consequently,
\begin{equation}
G^{\sigma}=\frac{1}{O}\sum_Iz_IG_I^{\sigma},
\quad
\bar x_{\gamma}=-\sqrt{\Delta\tau}
\Tr\!\left[L^{\gamma}\sum_{\sigma}G^{\sigma}\right].
\label{eq:si-native-density-force}
\end{equation}

For later comparison, introduce determinant-resolved Cholesky contractions
\begin{equation}
t_{I\gamma}^{\sigma}=\Tr(L^{\gamma}G_I^{\sigma}),
\quad
x_{I\gamma}^{\sigma}
=\Tr(L^{\gamma}G_I^{\sigma}L^{\gamma}G_I^{\sigma}).
\label{eq:si-native-chol-features}
\end{equation}
The one- and two-body parts of the determinant local energy are
\begin{align}
E_{1,I}&=\sum_{\sigma}\Tr(h^{\sigma}G_I^{\sigma}),\\
E_{2,I}&=\frac{1}{2}\sum_{\gamma}
\left[
\left(t_{I\gamma}^{\alpha}+t_{I\gamma}^{\beta}\right)^2
-x_{I\gamma}^{\alpha}-x_{I\gamma}^{\beta}
\right].
\label{eq:si-native-energies}
\end{align}
The exact multideterminant components are
\begin{equation}
E_1=\frac{\sum_Iz_IE_{1,I}}{O},
\quad
E_2=\frac{\sum_Iz_IE_{2,I}}{O},
\quad
E_L=E_0+E_1+E_2.
\label{eq:si-native-msd-energy}
\end{equation}
The two-body average in Eq.~\eqref{eq:si-native-msd-energy} must retain the
determinant-resolved quadratic functional. In general,
\begin{equation}
E_2\!\left[\frac{\sum_Iz_IG_I^{\alpha}}{O},
\frac{\sum_Iz_IG_I^{\beta}}{O}\right]
\ne \frac{1}{O}\sum_Iz_IE_2[G_I^{\alpha},G_I^{\beta}],
\label{eq:si-wrong-average}
\end{equation}
because the expression on the left introduces cross terms between different CI configurations. All accelerated backends in the main text preserve the right-hand side of Eq.~\eqref{eq:si-wrong-average} exactly.

\section{Reference determinant and particle--hole overlap ratios}
\label{sec:si-reference}

We use an orbital basis in which every trial determinant is an occupation selector. For spin $\sigma$, choose a reference occupation
\begin{equation}
\cO_0^{\sigma}=(o_1,\ldots,o_{N_{\sigma}})
\end{equation}
and define
\begin{equation}
B_0^{\sigma}=\Phi^{\sigma}[\cO_0^{\sigma},:],
\quad
O_0^{\sigma}=\det B_0^{\sigma},
\quad
K^{\sigma}=\Phi^{\sigma}(B_0^{\sigma})^{-1}.
\label{eq:si-reference-kernel}
\end{equation}
The inverse in Eq.~\eqref{eq:si-reference-kernel} is implemented through a linear solve. The rows of $K^{\sigma}$ span the full $M$-orbital space and its columns label positions in the reference occupation. In particular,
\begin{equation}
K^{\sigma}[\cO_0^{\sigma},:]=I_{N_{\sigma}}.
\end{equation}

Let $E_0^{\sigma\dagger}$ be the row-selector matrix associated with
$\cO_0^{\sigma}$. The reference transition density and its complementary
projector are
\begin{equation}
G_0^{\sigma}=K^{\sigma}E_0^{\sigma\dagger},
\quad
Q_0^{\sigma}=I-G_0^{\sigma}.
\label{eq:si-reference-projectors}
\end{equation}
Thus, the columns of $G_0^{\sigma}$ at the reference occupied indices equal the columns of $K^{\sigma}$ and all other columns vanish. The transition density is generally non-Hermitian, but it is idempotent: $(G_0^{\sigma})^2=G_0^{\sigma}$.

A target spin determinant $a$ is obtained by replacing the occupied positions
\begin{equation}
\vct h_a^{\sigma}=(h_{a1},\ldots,h_{ak})
\end{equation}
with particle orbitals
\begin{equation}
\vct p_a^{\sigma}=(p_{a1},\ldots,p_{ak}).
\end{equation}
Here $h_{a\nu}$ is a column of $K^{\sigma}$, or equivalently a position in $\cO_0^{\sigma}$, whereas $p_{a\mu}$ is a full-space orbital index. A stored phase $s_a^{\sigma}\in\{+1,-1\}$ maps the aligned row-replacement ordering to the canonical ordering of the target occupation. Define the small excitation matrix
\begin{equation}
(A_a^{\sigma})_{\mu\nu}
=K^{\sigma}_{p_{a\mu},h_{a\nu}}.
\label{eq:si-excitation-matrix}
\end{equation}
The determinant row-replacement lemma then gives
\begin{equation}
r_a^{\sigma}
\equiv
\frac{\langle D_a^{\sigma}|\Phi^{\sigma}\rangle}
{\langle D_0^{\sigma}|\Phi^{\sigma}\rangle}
=s_a^{\sigma}\det A_a^{\sigma}.
\label{eq:si-overlap-ratio}
\end{equation}
Only the $N_{\sigma}\times N_{\sigma}$ reference matrix is factorized at the full occupied-space scale. Every determinant-dependent overlap is reduced to a determinant whose dimension is the excitation rank $k$.

For a full configuration $I$, let $a(I)$ and $b(I)$ denote its alpha and beta spin strings. After factoring out the two reference overlaps,
\begin{align}
O&=O_0^{\alpha}O_0^{\beta}\cS,\\
\cS&=\sum_{I=1}^{\Nc}c_I^*
r_{a(I)}^{\alpha}r_{b(I)}^{\beta}.
\label{eq:si-relative-overlap}
\end{align}

\section{Determinant derivatives and cofactor conventions}
\label{sec:si-cofactors}

The GWT insertions needed below can be expressed as derivatives of the excitation determinant. For a $k\times k$ matrix $A$, define
\begin{align}
\frac{\partial\det A}{\partial A_{\mu\nu}}
&=\adj(A)_{\nu\mu}
=(-1)^{\mu+\nu}\det A[\widehat\mu|\widehat\nu],
\label{eq:si-first-derivative}\\
\frac{\partial^2\det A}
{\partial A_{\mu\nu}\partial A_{\rho\lambda}}
&=\epsilon_{\mu\nu\rho\lambda}
\det A[\widehat{\mu\rho}|\widehat{\nu\lambda}],
\label{eq:si-second-derivative}
\end{align}
where a hat denotes a deleted row or column. For distinct rows and columns,
\begin{equation}
\epsilon_{\mu\nu\rho\lambda}
=(-1)^{\mu+\nu+\rho+\lambda}
\operatorname{sgn}(\rho-\mu)\operatorname{sgn}(\lambda-\nu),
\label{eq:si-second-sign}
\end{equation}
and the second derivative vanishes if $\rho=\mu$ or $\lambda=\nu$. These relations are the determinant form of the off-diagonal or generalized Wick construction~\cite{Wick1950,Lowdin1955,Thouless1960,BalianBrezin1969, Burton2021}.

Including the fermionic phase, define the first- and second-derivative objects for spin string $a$ by
\begin{align}
D_{a,\nu\mu}^{(1),\sigma}
&=s_a^{\sigma}\adj(A_a^{\sigma})_{\nu\mu},
\label{eq:si-D1}\\
D_{a,\mu\nu\rho\lambda}^{(2),\sigma}
&=s_a^{\sigma}
\frac{\partial^2\det A_a^{\sigma}}
{\partial(A_a^{\sigma})_{\mu\nu}
 \partial(A_a^{\sigma})_{\rho\lambda}}.
\label{eq:si-D2}
\end{align}
For arbitrary $k\times k$ directions $U$ and $V$, the directional
derivatives are consequently
\begin{align}
\mathrm D r_a^{\sigma}[U]
&=\sum_{\mu\nu}D_{a,\nu\mu}^{(1),\sigma}U_{\mu\nu},
\label{eq:si-first-directional}\\
\mathrm D^2 r_a^{\sigma}[U,V]
&=\sum_{\mu\nu\rho\lambda}
D_{a,\mu\nu\rho\lambda}^{(2),\sigma}
U_{\mu\nu}V_{\rho\lambda}.
\label{eq:si-second-directional}
\end{align}

Equations~\eqref{eq:si-first-derivative}--\eqref{eq:si-second-directional} are polynomial identities. They remain valid when $A_a^{\sigma}$ is singular and should not be implemented through $\adj(A)=\det(A)A^{-1}$. The latter form both fails at exact singularity and can lose accuracy when the excitation-overlap ratio is small. The reference matrix $B_0^{\sigma}$ must still be sufficiently nonsingular to construct $K^{\sigma}$.

The rank-zero reference string is assigned
\begin{equation}
r_0^{\sigma}=1,
\quad
D_0^{(1),\sigma}=0,
\quad
D_0^{(2),\sigma}=0.
\label{eq:si-rank-zero}
\end{equation}
For rank one, the first cofactor is unity before multiplication by the stored phase, and all second derivatives vanish. These conventions remove special cases from the estimator assembly.

\section{One-body GWT identity and the mixed density}
\label{sec:si-one-body}

The central unnormalized one-body identity is
\begin{equation}
r_a^{\sigma}G_a^{\sigma}
=r_a^{\sigma}G_0^{\sigma}+\Delta_a^{\sigma},
\label{eq:si-unnormalized-density}
\end{equation}
where
\begin{equation}
\Delta_a^{\sigma}
=K^{\sigma}[:,\vct h_a^{\sigma}]
D_a^{(1),\sigma}
Q_0^{\sigma}[\vct p_a^{\sigma},:].
\label{eq:si-density-correction}
\end{equation}
In Eq.~\eqref{eq:si-density-correction}, $D_a^{(1),\sigma}$ has its first index in the local hole space and its second index in the local particle space. Equation~\eqref{eq:si-unnormalized-density} follows either from the off-diagonal Wick theorem or directly by differentiating the row-replacement determinant. Unlike a formula for $G_a^{\sigma}$ itself, its two sides remain finite when $r_a^{\sigma}=0$.

For an arbitrary one-body matrix $X^{\sigma}$, define
\begin{align}
x_0^{\sigma}[X]
&=\Tr(X^{\sigma}G_0^{\sigma}),\\
\mathcal B_X^{\sigma}
&=Q_0^{\sigma}X^{\sigma}K^{\sigma},
\label{eq:si-general-one-body-direction}\\
\ell_a^{\sigma}[X]
&=\sum_{\mu\nu}D_{a,\nu\mu}^{(1),\sigma}
(\mathcal B_X^{\sigma})_{p_{a\mu},h_{a\nu}}.
\label{eq:si-general-one-body-feature}
\end{align}
Taking the trace of Eq.~\eqref{eq:si-unnormalized-density} gives
\begin{equation}
r_a^{\sigma}\Tr(X^{\sigma}G_a^{\sigma})
=r_a^{\sigma}x_0^{\sigma}[X]+\ell_a^{\sigma}[X].
\label{eq:si-one-body-insertion}
\end{equation}
Thus, a one-body insertion is a reference contraction plus a first directional derivative of the excitation determinant.

To assemble the multideterminant density, embed the local first cofactors into the common reference-hole by full-particle space:
\begin{align}
\cC_{hp}^{\alpha}
&=\sum_Ic_I^*r_{b(I)}^{\beta}
\left[D_{a(I)}^{(1),\alpha}\right]_{hp}^{\mathrm{embedded}},\\
\cC_{hp}^{\beta}
&=\sum_Ic_I^*r_{a(I)}^{\alpha}
\left[D_{b(I)}^{(1),\beta}\right]_{hp}^{\mathrm{embedded}}.
\label{eq:si-compact-cofactor}
\end{align}
The embedding places a local element $(\nu,\mu)$ at $(h_{a\nu},p_{a\mu})$ and sets all other elements to zero. Summing Eq.~\eqref{eq:si-unnormalized-density} over configurations then gives
\begin{equation}
G^{\sigma}
=G_0^{\sigma}
+\frac{K^{\sigma}\cC^{\sigma}Q_0^{\sigma}}{\cS}.
\label{eq:si-gwt-green}
\end{equation}
The expensive full-$M$ matrix multiplications in Eq.~\eqref{eq:si-gwt-green} are performed only after the CI sum has reduced all first cofactors to $\cC^{\sigma}$. The force bias follows immediately from Eqs.~\eqref{eq:si-native-density-force} and \eqref{eq:si-gwt-green}.

\section{Two-body GWT identities}
\label{sec:si-two-body}

For each spin and Cholesky vector, precompute the reference quantities
\begin{align}
t_{0\gamma}^{\sigma}
&=\Tr(L^{\gamma}G_0^{\sigma}),\\
x_{0\gamma}^{\sigma}
&=\Tr(L^{\gamma}G_0^{\sigma}L^{\gamma}G_0^{\sigma}),\\
B^{\sigma\gamma}
&=Q_0^{\sigma}L^{\gamma}K^{\sigma},
\label{eq:si-B-direction}\\
C^{\sigma\gamma}
&=Q_0^{\sigma}L^{\gamma}G_0^{\sigma}L^{\gamma}K^{\sigma}.
\label{eq:si-C-direction}
\end{align}
Only the rows and columns selected by a given excitation are used in its small-space contractions. Define the unnormalized connected features
\begin{align}
f_{a\gamma}^{\sigma}
&=\sum_{\mu\nu}D_{a,\nu\mu}^{(1),\sigma}
B_{p_{a\mu},h_{a\nu}}^{\sigma\gamma},
\label{eq:si-f-feature}\\
g_{a\gamma}^{\sigma}
&=\sum_{\mu\nu}D_{a,\nu\mu}^{(1),\sigma}
C_{p_{a\mu},h_{a\nu}}^{\sigma\gamma},
\label{eq:si-g-feature}\\
d_a^{\sigma}
&=\frac{1}{2}\sum_{\gamma\mu\nu\rho\lambda}
D_{a,\mu\nu\rho\lambda}^{(2),\sigma}
B_{p_{a\mu},h_{a\nu}}^{\sigma\gamma}
B_{p_{a\rho},h_{a\lambda}}^{\sigma\gamma}.
\label{eq:si-d-feature}
\end{align}
The first of these is the directional derivative $\mathrm D r_a^{\sigma}[B^{\sigma\gamma}]$. The last is one half of the second directional derivative, summed over the Cholesky directions.

For a single spin determinant and a fixed $\gamma$, the same-spin Wick
functional obeys
\begin{align}
&\frac{r_a^{\sigma}}{2}
\left[
(t_{a\gamma}^{\sigma})^2-x_{a\gamma}^{\sigma}
\right]
\nonumber\\
&\quad=
\frac{r_a^{\sigma}}{2}
\left[
(t_{0\gamma}^{\sigma})^2-x_{0\gamma}^{\sigma}
\right]
+t_{0\gamma}^{\sigma}f_{a\gamma}^{\sigma}
-g_{a\gamma}^{\sigma}
+\frac{1}{2}
\mathrm D^2r_a^{\sigma}
[B^{\sigma\gamma},B^{\sigma\gamma}].
\label{eq:si-same-spin-identity}
\end{align}
This identity is an unnormalized form of the same-spin generalized Wick theorem. When $r_a^{\sigma}\ne0$, it can also be verified by inserting Eq.~\eqref{eq:si-unnormalized-density} into the quadratic functional. The apparently rational terms cancel to the second determinant derivative. The polynomial form in Eq.~\eqref{eq:si-same-spin-identity} remains valid by continuity at $r_a^{\sigma}=0$.

The cross-spin direct term factorizes because alpha and beta operators act in different spin sectors. For a full configuration $(a,b)$,
\begin{align}
r_a^{\alpha}r_b^{\beta}
t_{a\gamma}^{\alpha}t_{b\gamma}^{\beta}
=& r_a^{\alpha}r_b^{\beta}
t_{0\gamma}^{\alpha}t_{0\gamma}^{\beta}
\nonumber\\
&+r_b^{\beta}t_{0\gamma}^{\beta}f_{a\gamma}^{\alpha}
+r_a^{\alpha}t_{0\gamma}^{\alpha}f_{b\gamma}^{\beta}
+f_{a\gamma}^{\alpha}f_{b\gamma}^{\beta}.
\label{eq:si-cross-spin-identity}
\end{align}
The last term is the nonseparable alpha--beta first-connected contribution. It is the origin of the exact $O(\Nc X)$ coefficient contraction retained by the differentiated-DAG algorithm.

\section{Complete flat GWT AFQMC kernel}
\label{sec:si-flat-kernel}

The reference one-body and two-body energies are
\begin{align}
E_{1,0}
&=\sum_{\sigma}\Tr(h^{\sigma}G_0^{\sigma}),
\label{eq:si-reference-e1}\\
E_{2,0}
&=\frac{1}{2}\sum_{\gamma}
\left[
(t_{0\gamma}^{\alpha}+t_{0\gamma}^{\beta})^2
-x_{0\gamma}^{\alpha}-x_{0\gamma}^{\beta}
\right].
\label{eq:si-reference-e2}
\end{align}
Define the one-body particle--hole intermediate and feature
\begin{align}
H^{\sigma}&=Q_0^{\sigma}h^{\sigma}K^{\sigma},
\label{eq:si-H-direction}\\
e_{1a}^{\sigma}
&=\sum_{\mu\nu}D_{a,\nu\mu}^{(1),\sigma}
H_{p_{a\mu},h_{a\nu}}^{\sigma}.
\label{eq:si-e1-feature}
\end{align}
Using Eq.~\eqref{eq:si-one-body-insertion}, the complete one-body local-energy component is
\begin{equation}
E_1=E_{1,0}+\frac{1}{\cS}\sum_Ic_I^*
\left[
r_{b(I)}^{\beta}e_{1a(I)}^{\alpha}
+r_{a(I)}^{\alpha}e_{1b(I)}^{\beta}
\right].
\label{eq:si-flat-e1}
\end{equation}

Let
\begin{equation}
T_{0\gamma}=t_{0\gamma}^{\alpha}+t_{0\gamma}^{\beta}
\label{eq:si-T0}
\end{equation}
and define the spin-string two-body correction
\begin{equation}
u_a^{\sigma}
=\sum_{\gamma}
\left(T_{0\gamma}f_{a\gamma}^{\sigma}
-g_{a\gamma}^{\sigma}\right)+d_a^{\sigma}.
\label{eq:si-u-feature}
\end{equation}
Combining the same-spin identity in Eq.~\eqref{eq:si-same-spin-identity} with the opposite-spin identity in Eq.~\eqref{eq:si-cross-spin-identity}
gives
\begin{align}
E_2=E_{2,0}+\frac{1}{\cS}\sum_Ic_I^*\bigg[
&r_{b(I)}^{\beta}u_{a(I)}^{\alpha}
+r_{a(I)}^{\alpha}u_{b(I)}^{\beta}
\nonumber\\
&+\sum_{\gamma}f_{a(I)\gamma}^{\alpha}
f_{b(I)\gamma}^{\beta}
\bigg].
\label{eq:si-flat-e2}
\end{align}
Equations~\eqref{eq:si-relative-overlap}, \eqref{eq:si-gwt-green}, \eqref{eq:si-native-density-force}, \eqref{eq:si-flat-e1}, and \eqref{eq:si-flat-e2} constitute the complete flat GWT AFQMC kernel used as a baseline in the main text.

It is useful to emphasize that $e_{1a}^{\sigma}$, $f_{a\gamma}^{\sigma}$, $g_{a\gamma}^{\sigma}$, and $d_a^{\sigma}$ are \emph{unnormalized} connected features. None is divided by $r_a^{\sigma}$. This choice avoids numerical singularities for individual zero- or near-zero-overlap configurations. Normalization occurs only once, through the total relative overlap $\cS$.

\section{Unique-spin environments}
\label{sec:si-environments}

The derivation above is written as a flat loop over full configurations. For the DAG formulations, repeated spin strings are first deduplicated. Let $a=1,\ldots,\Sa$ and $b=1,\ldots,\Sb$ enumerate the unique alpha and beta strings. The opposite-spin dependence of a fixed string is collected into the environment weights
\begin{align}
q_a^{\alpha}
&=\sum_{I:a(I)=a}c_I^*r_{b(I)}^{\beta},\\
q_b^{\beta}
&=\sum_{I:b(I)=b}c_I^*r_{a(I)}^{\alpha}.
\label{eq:si-environments}
\end{align}
They satisfy
\begin{equation}
\cS=\sum_{a=1}^{\Sa}q_a^{\alpha}r_a^{\alpha}
=\sum_{b=1}^{\Sb}q_b^{\beta}r_b^{\beta}.
\label{eq:si-environment-overlap}
\end{equation}
Every term involving a nontrivial feature in only one spin sector can be reduced using $q_a^{\sigma}$. For example,
\begin{equation}
\sum_Ic_I^*r_{b(I)}^{\beta}e_{1a(I)}^{\alpha}
=\sum_{a=1}^{\Sa}q_a^{\alpha}e_{1a}^{\alpha}.
\label{eq:si-environment-example}
\end{equation}
The last term of Eq.~\eqref{eq:si-flat-e2} contains nontrivial features from both spin sectors and cannot in general be reduced to independent environment sums. It remains
\begin{equation}
\sum_{I\gamma}c_I^*f_{a(I)\gamma}^{\alpha}
f_{b(I)\gamma}^{\beta}.
\label{eq:si-irreducible-cross}
\end{equation}
Equation~\eqref{eq:si-irreducible-cross} preserves the original sparse CI coefficient coupling exactly.

\section{Algebraic equivalence to the two DAG formulations}
\label{sec:si-dag-equivalence}

The flat GWT evaluates the objects in Eqs.~\eqref{eq:si-overlap-ratio}, \eqref{eq:si-D1}, and \eqref{eq:si-D2} separately. The explicit-cofactor DAG represents each deleted minor as a shared node. For canonically ordered row and column tuples $R$ and $C$, define
\begin{equation}
M^{\sigma}(R,C)=\det K^{\sigma}[R,C],
\quad
M^{\sigma}(\varnothing,\varnothing)=1.
\label{eq:si-minor-node}
\end{equation}
A first-row Laplace expansion gives the arithmetic recurrence
\begin{equation}
M(R,C)=\sum_{j=1}^{|C|}(-1)^{j-1}K_{r_1c_j}
M(R\setminus r_1,C\setminus c_j).
\label{eq:si-minor-recurrence}
\end{equation}
The explicit-cofactor DAG compiles the top-level excitation minors and all single- and double-deletion roots requested by the flat formulas. It changes neither the signs nor the contractions. It only merges identical minor evaluations.

The differentiated DAG compiles only the top-level overlap roots and their recursive dependencies. For spin $\sigma$, define the environment-weighted scalar
\begin{equation}
F^{\sigma}(K^{\sigma})
=\sum_{a=1}^{S_{\sigma}}q_a^{\sigma}s_a^{\sigma}
M^{\sigma}(P_a^{\sigma},H_a^{\sigma}).
\label{eq:si-weighted-scalar}
\end{equation}
Let $v_n$ be the value of graph node $n$. For an edge $e=(n\rightarrow m)$, let $s_e$ be its Laplace sign and $(r_e,c_e)$ the associated element of $K^\sigma$. Then
\begin{equation}
v_n=\sum_{e=(n\rightarrow m)}s_eK^\sigma_{r_ec_e}v_m.
\label{eq:si-node-recurrence}
\end{equation}
For the adjoints $\bar v_n=\partial F^\sigma/\partial v_n$, the top-level roots are seeded by $\bar v_{n(a)}\mathrel{+}=q_a^\sigma s_a^\sigma$. A reverse topological traversal applies
\begin{align}
\Gamma^\sigma_{r_ec_e}&\mathrel{+}=\bar v_ns_ev_m,\\
\bar v_m&\mathrel{+}=\bar v_ns_eK^\sigma_{r_ec_e}.
\label{eq:si-reverse-recurrence}
\end{align}
Adjoints arriving from all parents are summed, so each shared subexpression is differentiated once.
With the opposite-spin environments held fixed during differentiation,
$F^{\sigma}=\cS$. Its gradient is
\begin{equation}
\Gamma_{ph}^{\sigma}
=\frac{\partial F^{\sigma}}{\partial K_{ph}^{\sigma}}
=\sum_aq_a^{\sigma}
\left[D_a^{(1),\sigma}\right]_{hp}^{\mathrm{embedded}}.
\label{eq:si-aggregate-gradient}
\end{equation}
Therefore,
\begin{equation}
\cC^{\sigma}=(\Gamma^{\sigma})^T,
\label{eq:si-C-gradient}
\end{equation}
and a single reverse traversal of the shared overlap graph produces exactly the cofactor sum required in Eq.~\eqref{eq:si-gwt-green}. The same aggregate gradient gives
\begin{align}
\Gamma^{\sigma}:H^{\sigma}
&=\sum_aq_a^{\sigma}e_{1a}^{\sigma},\\
\Gamma^{\sigma}:B^{\sigma\gamma}
&=\sum_aq_a^{\sigma}f_{a\gamma}^{\sigma},\\
\Gamma^{\sigma}:C^{\sigma\gamma}
&=\sum_aq_a^{\sigma}g_{a\gamma}^{\sigma}.
\label{eq:si-gradient-contractions}
\end{align}

For a Cholesky direction, set
\begin{equation}
K^{\sigma}(t)=K^{\sigma}+tB^{\sigma\gamma}.
\end{equation}
If $v_n$, $\dot v_n^{\gamma}$, and $\ddot v_n^{\gamma}$ denote a graph-node value and its first and second derivatives with respect to $t$, differentiating Eq.~\eqref{eq:si-minor-recurrence} gives
\begin{align}
\dot v_n^{\gamma}
&=\sum_{e=(n\rightarrow m)}s_e
\left(B_{r_ec_e}^{\sigma\gamma}v_m
+K_{r_ec_e}^{\sigma}\dot v_m^{\gamma}\right),
\label{eq:si-first-tangent}\\
\ddot v_n^{\gamma}
&=\sum_{e=(n\rightarrow m)}s_e
\left(2B_{r_ec_e}^{\sigma\gamma}\dot v_m^{\gamma}
+K_{r_ec_e}^{\sigma}\ddot v_m^{\gamma}\right).
\label{eq:si-second-tangent}
\end{align}
At a top-level root $a$,
\begin{equation}
f_{a\gamma}^{\sigma}=s_a^{\sigma}\dot v_a^{\gamma},
\quad
\sum_aq_a^{\sigma}d_a^{\sigma}
=\frac{1}{2}\sum_{a\gamma}q_a^{\sigma}s_a^{\sigma}
\ddot v_a^{\gamma}.
\label{eq:si-root-tangents}
\end{equation}
Thus, Eqs.~\eqref{eq:si-aggregate-gradient}--\eqref{eq:si-root-tangents} evaluate the same first- and second-cofactor contractions as the flat GWT, without constructing a separate $D_a^{(1),\sigma}$ or $D_a^{(2),\sigma}$ tensor for every spin string.

\begin{table}[htb]
\caption{One-to-one correspondence between the flat GWT objects and their graph evaluations. Every row is an algebraic identity, not an approximation.}
\label{tab:si-correspondence}
\centering
\footnotesize
\begin{tabular}{>{\raggedright\arraybackslash}p{0.25\linewidth}
               >{\raggedright\arraybackslash}p{0.31\linewidth}
               >{\raggedright\arraybackslash}p{0.32\linewidth}}
\toprule
Required quantity & Flat or explicit-cofactor form & Differentiated-DAG form\\
\midrule
Spin overlap ratio
& $s_a\det A_a$
& Signed top-level overlap node\\
\hline
Aggregate one-body cofactor
& $\sum_aq_aD_a^{(1)}$
& Reverse gradient $\Gamma=\partial F/\partial K$\\
\hline
One connected insertion
& $\sum_aq_aD_a^{(1)}:X$
& $\Gamma:X$\\
\hline
Per-string Cholesky feature
& $D_a^{(1)}:B^{\gamma}$
& Signed top-root tangent $s_a\dot v_a^{\gamma}$\\
\hline
Weighted same-spin double insertion
& $\frac12\sum_{a\gamma}q_aD_a^{(2)}:
(B^{\gamma}\otimes B^{\gamma})$
& $\frac12\sum_{a\gamma}q_as_a\ddot v_a^{\gamma}$\\
\bottomrule
\end{tabular}
\end{table}

\section{Operation counts and storage of the flat baseline}
\label{sec:si-cost}

Let $k_{I\sigma}$ be the excitation rank encountered for configuration $I$ and spin $\sigma$. If the flat implementation explicitly requests every nonzero derivative entry, its structural counts are
\begin{align}
N_{\mathrm{det}}
&=\sum_{I\sigma}\mathbf 1[k_{I\sigma}>0],\\
N_{\mathrm{first}}
&=\sum_{I\sigma}k_{I\sigma}^2,\\
N_{\mathrm{second}}
&=\sum_{I\sigma}k_{I\sigma}^2(k_{I\sigma}-1)^2.
\label{eq:si-flat-counts}
\end{align}
The second count treats the two differentiated matrix elements as ordered. Entries sharing a row or column are zero. Direct contraction over all Cholesky directions has a schematic determinant-dependent cost
\begin{equation}
T_{\mathrm{flat,small}} =O\!\left(\sum_{I\sigma}
[k_{I\sigma}^3+Xk_{I\sigma}^2+Xk_{I\sigma}^4]\right),
\label{eq:si-flat-cost}
\end{equation}
apart from the cost of evaluating the deleted minors themselves. For bounded excitation rank, this work is linear in $\Nc$, but its prefactor is dominated by the repeated second-cofactor contractions.

If every deleted minor is itself evaluated by a dense determinant call, the additional per-string arithmetic is schematically
\begin{equation}
O\!\left[
k^2(k-1)^3+k^2(k-1)^2\max(k-2,1)^3
\right],
\label{eq:si-deleted-minor-cost}
\end{equation}
for the first- and second-cofactor requests, respectively. The structural counts in Eq.~\eqref{eq:si-flat-counts} are more implementation independent: small fixed-rank determinants may be evaluated by specialized formulas, but the same derivative entries must still be produced and contracted in the flat organization.

The full-orbital intermediates $H^{\sigma}$, $B^{\sigma\gamma}$, and $C^{\sigma\gamma}$ are built only for the two reference spin determinants. Thus, flat GWT already removes the factor of $\Nc$ from the principal system-size contractions compared with na\"ive determinant summation, as in previous reference-based multideterminant formulations~\cite{MahajanSharma2020, MahajanSharma2021,MahajanLeeSharma2022,MahajanResponse2023}. Its remaining bottleneck is the repeated small determinant/cofactor algebra.

When configurations are streamed, the determinant-specific temporary storage is $O(k_{\max}^4+Xk_{\max})$ in addition to the reference intermediates and the two aggregate first-cofactor matrices. Storing derivative tensors for all configurations would instead require $O(\Nc k_{\max}^4)$ memory and is unnecessary. The explicit-cofactor DAG reduces repeated determinant-minor evaluation but still reconstructs the per-string derivative objects. The differentiated DAG removes those derivative-object storage and reconstruction costs by evaluating only the contractions listed in Table~\ref{tab:si-correspondence}.

Neither graph formulation changes the exact sparse coefficient contraction in Eq.~\eqref{eq:si-irreducible-cross}. Without additional structure in the CI coefficient matrix, this contribution is $O(\Nc X)$ and is the asymptotic boundary of the present exact differentiated-DAG algorithm.

\section{Controlled benchmark construction and extended results}
\label{sec:si-controlled-benchmark}

The controlled benchmark isolates determinant-count and excitation-graph effects from molecule-specific integral structure, trial-selection protocols, and Monte Carlo sampling noise. All five backends in these synthetic, or ``fake-molecule,'' comparisons are implemented in a common Python/NumPy prototype and are intended as proof-of-principle tests of their algebraic organization, rather than as cross-code production benchmarks. The main series uses $M=40$ spatial orbitals, $N_\alpha=N_\beta=10$, $X=100$ dense complex-Hermitian Cholesky matrices, and trial expansions with maximum excitation rank four. Reference rows of each walker contain an identity contribution plus complex Gaussian noise of standard scale 0.08, producing well-conditioned but nontrivial overlaps. The one-body matrices are dense complex Hermitian matrices scaled by $1/M$, and the Cholesky matrices are scaled by $0.2/\sqrt M$.

For each determinant count, the $\alpha$ pool contains $\Sa=\lceil\Nc^p\rceil$ unique strings and the $\beta$ pool contains $\Sb=\lceil\Nc/\Sa\rceil$, with $p=1/2$ in the balanced benchmark. Distinct $\alpha$--$\beta$ pairs are sampled without replacement. Nonreference strings are random particle--hole excitations whose rank is sampled uniformly from one through four. CI coefficients are complex Gaussian numbers, the reference coefficient is boosted, and the coefficient vector is normalized. Determinant counts are 30, 100, 300, 1000, 3000, and 10000, with an independent deterministic seed at each size.

All timings use a single Python 3.12.13 process with NumPy 2.3.5 and one BLAS thread on a MacBook Pro with a 14-core Apple M4 Pro processor and 48~GB of memory. Each value is the median of five evaluations after one untimed warmup. The timed region contains the overlap, both mixed Green's functions, force bias, $E_1$, and $E_2$, and excludes one-time trial preprocessing. Two structural profiles supplement the timing series: the unique-spin balance exponent is varied from $p=0.25$ to 0.75 at $\Nc=1000$, and the maximum excitation rank is varied from one to four at $\Nc=400$.

Figure~\ref{fig:si-validation}(a) reports the maximum relative error across the overlap, both spin Green's matrices, force bias, one-body energy, and two-body energy. Every accelerated backend agrees with na\"ive determinant summation to better than $1.2\times10^{-14}$. Panel (d) tests the intentionally incorrect replacement of the determinant-resolved two-body average by the single-determinant quadratic functional evaluated at the CI-averaged Green's function. The relative error ranges from $8.12\times10^{-3}$ at $\Nc=30$ to $4.21\times10^{-2}$ at $\Nc=10^4$ and reaches 0.334 at $\Nc=3000$. Agreement of the overlap and one-body quantities alone therefore does not validate the multideterminant local energy.

\begin{figure*}[htb]
\centering
\includegraphics[width=0.97\linewidth]{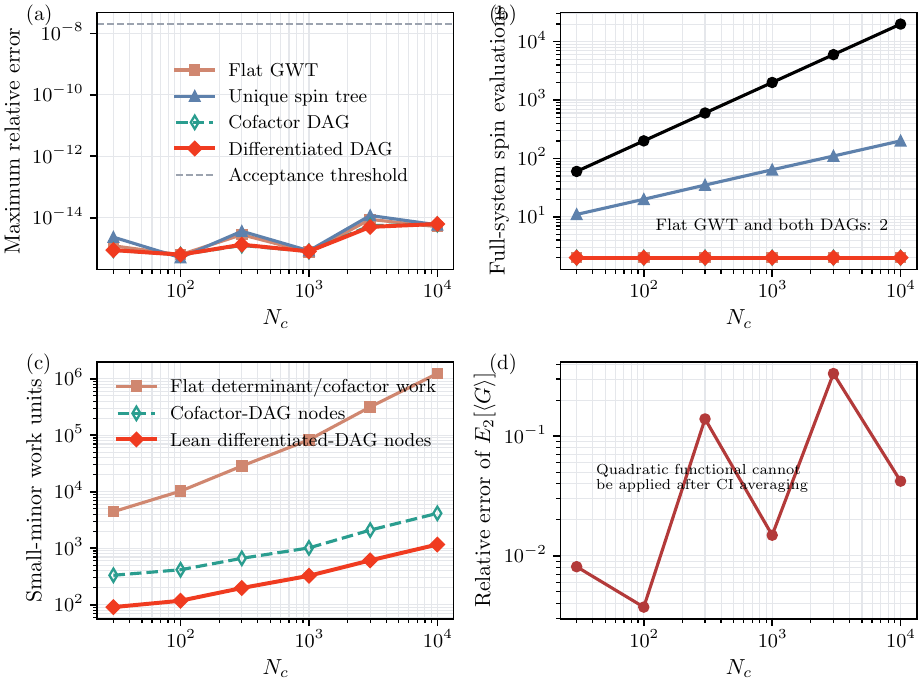}
\caption{Numerical equivalence and structural work in the controlled benchmark. (a) Maximum relative error of each accelerated backend against na\"ive determinant summation. (b) Full-system spin evaluations: the na\"ive count is $2\Nc$, the tree count is $\Sa+\Sb$, and flat GWT and both DAGs require two reference evaluations. (c) Flat determinant/cofactor requests compared with the node counts of the explicit-cofactor and differentiated DAGs. (d) Relative error produced by the incorrect substitution $\sum_I\omega_I E_2[G_I]\rightarrow E_2[\sum_I\omega_I G_I]$.}
\label{fig:si-validation}
\end{figure*}

Table~\ref{tab:si-timing} collects the largest controlled timing point. Although the flat GWT backend reduces full occupied-overlap factorizations from 20000 to two, its determinant-resolved cofactor loops dominate this small-$M$ prototype. This backend is an intentionally transparent, unbatched reference that evaluates the first- and second-cofactor expressions explicitly. Its timing isolates repeated cofactor work within the common prototype and should not be interpreted as a benchmark of optimized production GWT implementations, which employ compiled and batched kernels. Within the controlled prototype, sharing deleted minors in the explicit-cofactor DAG gives a 47.2-fold speedup over na\"ive summation, while differentiating the overlap graph raises the speedup to 1192. The difference isolates the benefit of moving derivative accumulation outside the unique-string loop.

\begin{table}[htb]
\caption{Largest controlled benchmark ($\Nc=10000$, $M=40$, $N_\alpha=N_\beta=10$, and $X=100$). The timing exponent uses all six determinant counts.}
\label{tab:si-timing}
\centering
\begin{tabular}{lrrrr}
\toprule
Method & Time (s) & Speedup & $p_{\rm time}$ & Max. rel. error \\
\midrule
Na\"ive & 150 & 1.00 & 1.00 & 0.00e+00 \\
Flat GWT & 317.9 & 0.47 & 1.02 & 5.13e-15 \\
Unique-spin tree & 1.533 & 97.82 & 0.50 & 5.90e-15 \\
Cofactor DAG & 3.174 & 47.24 & 0.50 & 6.08e-15 \\
Differentiated DAG & 0.1258 & 1192.45 & 0.28 & 6.31e-15 \\
\bottomrule
\end{tabular}
\end{table}

At $\Nc=10000$, flat GWT requests 19,800 top-level small determinants, 161,400 first-cofactor entries, and 1,059,200 second-cofactor entries. The explicit-cofactor graphs contain 4162 nodes and 8926 edges, whereas the differentiated overlap graphs contain 1167 nodes and 2460 edges. Relative to the 1,240,400 flat requests, the corresponding request-to-node ratios are 298 and 1063. These counts distinguish the savings from deleted-minor sharing and from eliminating derivative-only roots.

Figure~\ref{fig:si-structure} shows that compression is strongest near a balanced spin-string factorization and depends on excitation rank. At $\Nc=1000$, the $p=0.5$ trial represents 83.8 flat requests per explicit-cofactor node and 285.1 per differentiated DAG node. At maximum rank four, the differentiated graph has 245 nodes compared with 821 nodes in the cofactor graph. Its construction time is $6.36\times10^{-4}$~s, compared with $5.69\times10^{-3}$~s for the cofactor graph. In the largest timed case, total differentiated-DAG preprocessing takes 0.0125~s, less than one 0.1258~s walker evaluation, and is amortized over all walkers and time steps.

\begin{figure*}[htb]
\centering
\includegraphics[width=0.97\linewidth]{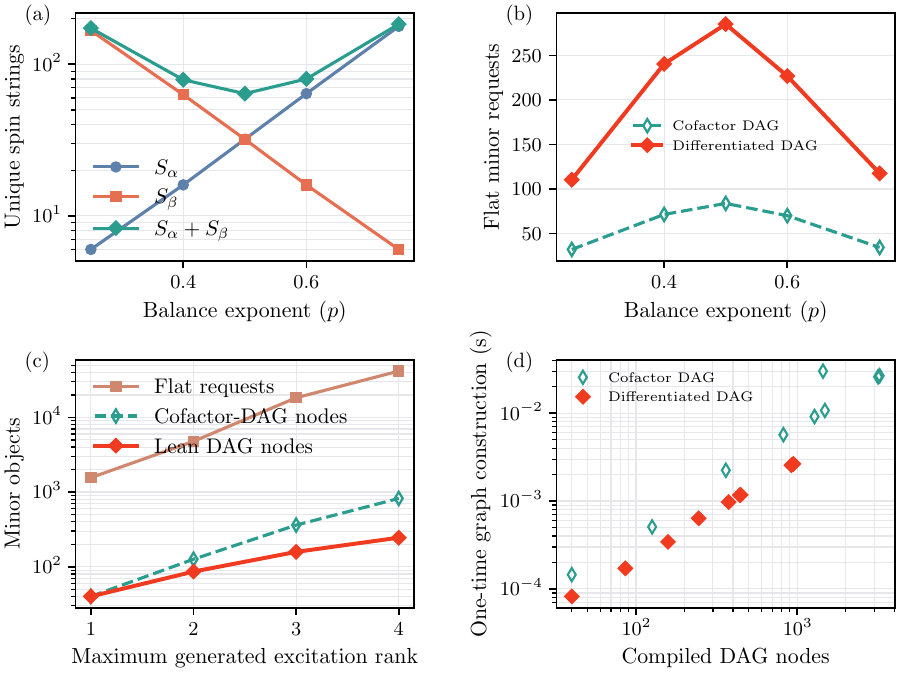}
\caption{Dependence of exact graph compression on trial structure. (a) Unique $\alpha$, $\beta$, and total spin strings as the balance exponent $p$ is varied at $\Nc=1000$. (b) Flat deleted-minor requests per cofactor- and differentiated-DAG node. (c) Flat requests and both graph sizes versus maximum generated excitation rank at $\Nc=400$. (d) One-time construction time versus compiled node count.}
\label{fig:si-structure}
\end{figure*}

The exact sparse opposite-spin contraction remains $O(\Nc X)$. Graph differentiation accelerates the shared polynomial response but cannot remove information carried by arbitrary independent CI coefficients. A low-rank or stochastic compression of this final contraction would be a separate, controlled approximation to the exact differentiated-DAG estimator.

\section{Cr$_2$ PEC with CASCI trial orbitals}
\label{sec:si-cr2-casci}

To isolate the effect of orbital optimization in the main-text Cr$_2$ calculation, Fig.~\ref{fig:cr2_pec_casci} reports the corresponding phaseless-AFQMC series obtained with unoptimized Dice SHCI-CASCI trial orbitals in the same CAS(12e,42o) active space. The determinant-count sequence extends from $10^3$ to $10^6$. All curves use the same minimum-alignment and shape-preserving interpolation protocol as the main-text CASSCF comparison.

\begin{figure*}[htb]
  \centering
  \includegraphics[width=\linewidth]{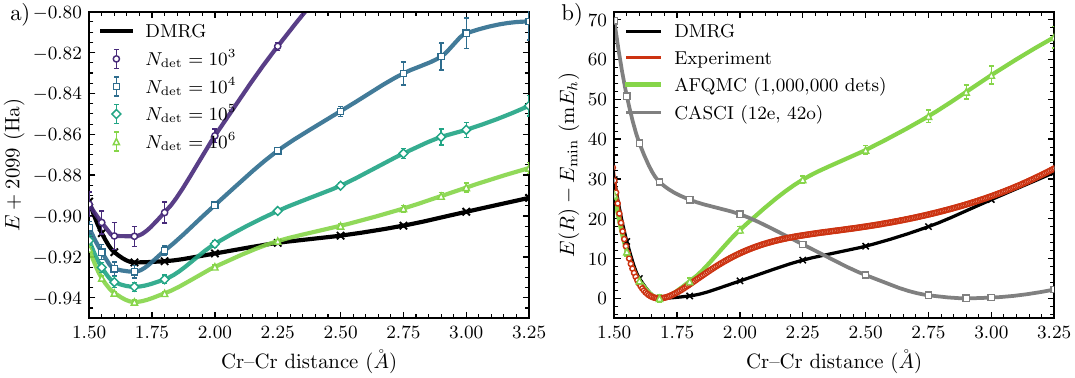}
  \caption{Similar as Figure~\ref{fig:cr2_pec} in main text, but with CASCI trial wavefunction.}
  \label{fig:cr2_pec_casci}
\end{figure*}

\end{widetext}

\end{document}